\documentclass[12pt,preprint]{aastex}

\newcommand{\nh}{\ensuremath{N_{\rm{H}}}}
\newcommand{\zfe}{\ensuremath{Z_{\rm{Fe}}}}
\newcommand{\lx}{\ensuremath{L_{\rm{X}}}}

\newcommand{\pwFP}{\ensuremath{\Gamma}}

\slugcomment{}
\shorttitle{New Evidence for High Activity of Sagittarius A*}
\shortauthors{M.~Nobukawa et al.}

\begin{document}
\title{New Evidence for High Activity of the Super-massive Black Hole in our Galaxy}
\author{
M.~Nobukawa,
S.~G.~Ryu,
T.~G.~Tsuru, 
and
K.~Koyama
}
\affil{Department of Physics, Graduate School of Science, Kyoto University, Sakyo-ku, Kyoto 606-8502, Japan}
\email{nobukawa@cr.scphys.kyoto-u.ac.jp}


\begin{abstract}
Prominent K-shell emission lines of neutral iron (hereafter, \ion{Fe}{1}-K lines) and hard-continuum 
X-rays from molecular clouds (MCs) in the Sagittarius B (Sgr B) region 
were found in 
two separate \textit{Suzaku} observations in 2005 and 2009. The X-ray flux of the 
\ion{Fe}{1}-K lines
decreased in correlation with the hard-continuum flux by a factor of 0.4--0.5 in four 
years, 
which is almost equal to the light-traveling across the MCs. The rapid and correlated 
time-variability, the equivalent width of the \ion{Fe}{1}-K lines, 
and the K-edge absorption depth of \ion{Fe}{1} are 
consistently explained by ''X-ray echoes'' due to the fluorescent and 
Thomson-scattering of an X-ray flare from an external source. 
The required flux of the X-ray flare depends on the distance to the MCs and its time duration. 
Even for a case with a minimum distance,
the flux is larger than those of the brightest Galactic X-ray sources. Based on these facts, 
we conclude that the super-massive black hole Sgr A* exhibited a large-flare  
a few hundred 
years ago with a luminosity of more than $4\times 10^{39}$ erg~s$^{-1}$ .
The ''X-ray echo'' from Sgr B, located a few hundred light-years from Sgr A*, 
has now reached at the Earth. 
\end{abstract}


\section{Introduction} 
One of the most significant topics of research is X-rays emission in the
Galactic center (GC) region, particularly the K-shell 
emission lines of neutral iron (hereafter, the \ion{Fe}{1}-K lines) from 
the giant molecular clouds 
(MCs: \citealt{Tsu99}) in Sagittarius B (Sgr B), located at $l=0.7$\arcdeg\ 
on the Galactic plane. 
X-rays and the \ion{Fe}{1}~K lines from Sgr B were first detected by 
the $Granat$/ART-P telescope (\citealt{Su93})
and the ASCA satellite (\citealt{Ko96}), respectively.
The MCs are extremely cold, 
and hence cannot self-excite neutral irons to a higher level in order to emit 
the \ion{Fe}{1}-K lines. 
The origin of \ion{Fe}{1}-K lines has been a mystery. From laboratory experiments, 
two plausible origins have been proposed: the fluorescence after inner-shell ionization 
by the injection of electrons or emission of X-rays with energies greater than 7.1 keV (the minimum ionization 
energy of \ion{Fe}{1}). The X-ray spectra of the MCs, which show a strong absorption edge at 7.1 keV 
together with intense \ion{Fe}{1}-K$\alpha$ (L- to K-shell transition) and \ion{Fe}{1}-K$\beta$ 
(M- to K-shell transition) lines, support the X-ray origin rather than the electron origin
(\citealt{Ko96, Mu00}).

The X-ray origin was established by the recent discovery of rapid (order of several years) 
time-variability of \ion{Fe}{1}-K lines \citep{Ko08, Inui09}, 
and from the fact that only photons
can produce such rapid variability in the MCs with size of the order of 10 
light-years. 
Increase in the rapid time variability of the \ion{Fe}{1}-K lines 
from other MCs near the GC
has been discovered \citep{Po10}. 
This suggests that the MCs have only been illuminated by an external X-ray source.
Furthermore, time variability in the continuum X-ray was found by Muno et al. (2007). 
In the X-ray fluorescent origin, Thomson scattered X-rays are also predicted 
in the hard X-ray band. 
In fact, 
the X-rays detected by \cite{Su93} were the Thomson scattered component.
\cite{Te10} discovered time variable hard X-rays (20--60 keV)
originating from the Sgr~B2 region.
In addition, Muno et al. (2007) found flickering continuum X-ray emissions 
from small clouds near the GC.
Accordingly, X-rays from the clouds may be called ''X-ray echoes.'' 
\ion{Fe}{1}-K lines and hard continuum X-rays are due to fluorescence and the 
Thomson scattering of external X-rays, respectively \citep{Ko96, Mu00, Te10}.
The ''X-ray echoes'' can easily produce apparent superluminal propagation,
as Ponti et al. (2010) found in the MCs near the GC (Sunyaev \& Churazov 1998).

The X-ray luminosity of these clouds is $\sim 10^{35}$~erg~s$^{-1}$, and 
hence an illuminating
object should be much brighter ($> 10^{39}$~erg~s$^{-1}$; Sunyaev \& Churazov 1998; 
Murakami et al. 2000). 
One proposed scenario is that Sgr A*, the super-massive black hole with around four million solar 
masses in the GC located at a distance of several hundred light-years from the MCs 
\citep{Ghez08}, 
was bright several hundred years ago with an X-ray luminosity millions times higher than the present 
luminosity of 10$^{33}$--10$^{34}$~erg~s$^{-1}$ \citep{Ba01}. 
The increase and decrease in the X-ray emission from the MCs may suggest 
intermittent activities of Sgr~A* several hundred years ago 
(Inui et al. 2009; Ponti et al. 2010).

However, because the required luminosity of the external X-ray source depends on the 
distance to the MCs, the duration of X-ray irradiation, and the size and density 
of the MCs, the currently available data does not exclude the possibility of intense X-ray flares 
from near-by transient sources. Recently, Ponti et al. (2010) discovered  
superluminal motion in the \ion{Fe}{1}-K emission from the MCs near the 
GC $(l,\ b)\sim(0^{\circ}.1,\ 0^{\circ}.1)$ and estimated that a large luminosity 
of $>1.3\times10^{38}$~erg~s$^{-1}$, which is almost
equal to the Eddington luminosity for a stellar mass object was required for 
illumination. They concluded that the past flare of Sgr A* had illuminated the MCs.
However, for other MCs emitting \ion{Fe}{1}~K, particularly, those in 
Sgr~B, there is no clear
evidence that Sgr A* was the only candidate for their illumination source. 
This paper provides clear evidence for the scenario of a past large-flare of Sgr A*. 

The errors in the estimation in this paper are at the 1$\sigma$ confidence level. The distance to the GC is 
assumed to be the canonical value of 26,000 light-years (\citealt{Ei03, Ghez08}). 
Because we use the Galactic coordinate system, 
we define ''east'' and ''west'' as the positive and 
negative Galactic longitude sides, respectively.

\section{Observation and Data Reduction}  
\label{observations}  
To identify the external X-ray source for the ''X-ray echo,'' Sgr B2 would be 
the best target because (i) it is one of the brightest X-ray echo sources, 
(ii) its three-dimensional position is well determined \citep{Reid09, Ryu09}, 
and (iii) the Galactic center 
diffuse X-ray emission (GCDX: \citealt{Ko96}) is relatively small and uniform. 
Therefore, possible contamination from the GCDX can be reliably subtracted.

We observed the Sgr B region with the X-ray CCD cameras 
the X-ray Imaging Spectrometer (XIS;  
\citealt{Ko07a}) aboard the Suzaku satellite \citep{Mi07}. To examine time variability, 
the observations were performed twice in October 2005 and September 2009, at 
identical pointing positions within 4\arcsec. 
The sequence numbers for the two observations are 100037060 and 504004020.
The effective exposure times in 2005 and 2009 were 21.3~h 
and 56.1~h, respectively. Four XIS (XIS0, 1, 2, and 3) and three XIS (XIS0, 1, and 3) 
cameras were available in 2005 and 2009, respectively. We discarded 
the data obtained in a one-fourth field (segment A) of XIS0 for both observations 
because it has become nonfunctional since June 2008.

Although the effective energy range of XIS is 0.2--10 keV, we used 
the 4--10 keV band data for the spectral analysis. 
This is because soft X-rays from the target (Sgr~B) are absorbed by 
the interstellar medium toward the GC region.
We obtained the data for non-X-ray background (NXB) 
caused by the cosmic-rays using \textit{xisnxbgen} (\citealt{Ta08}) and then subtracted 
the NXB from the observed data. 
We then corrected the local differences of the effective area for the data analysis 
(the vignetting correction). 

\section{Analysis} 
\label{analysis}
\subsection{X-ray Images} 
\begin{figure}[tb]
 \epsscale{1.0}
 \plotone{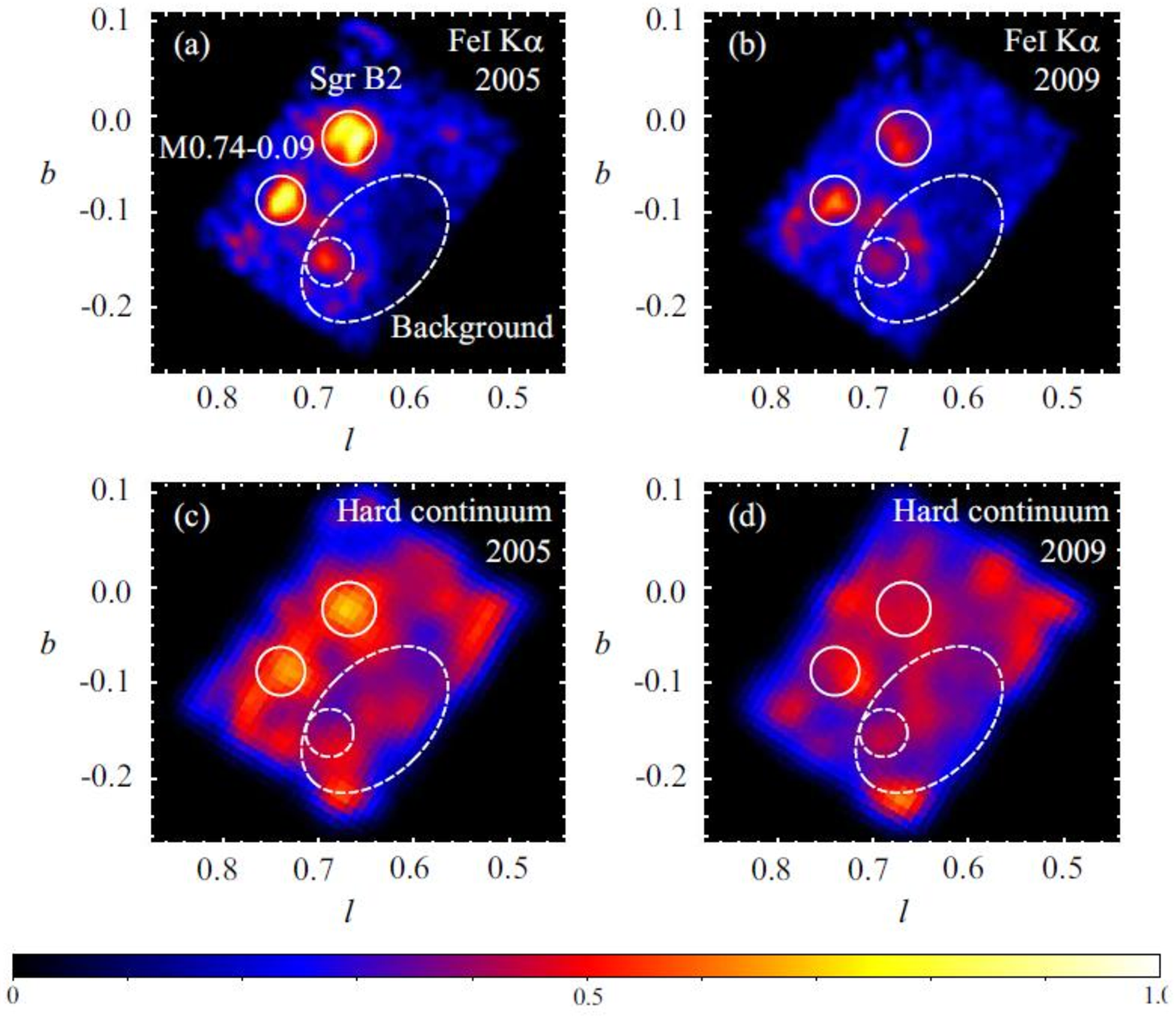}
 \caption{
 X-ray images for Sgr B2 and M\,0.74$-$0.09 in the field of view of 17.8\arcmin$\times$13.4\arcmin. 
 The vertical and horizontal axes are the Galactic longitude ($l$) and latitude ($b$) in degrees 
 (\arcdeg), respectively. (a): The \ion{Fe}{1}-K$\alpha$ (6.4 keV) band images in 2005. 
 The bin size is 8.3\arcsec\ and is smoothed with a Gaussian kernel of 1$\sigma$$=$42\arcsec. 
 (b): The same as (a) but in 2009. (c): The hard continuum (8-10 keV) band images in 2005. 
 The bin size is 33.3 \arcsec\ and is smoothed with a Gaussian kernel of 1$\sigma=$100\arcsec. 
 (d) The same as (c) but in 2009. The source regions for Sgr B2 and M\,0.74$-$0.09 are shown in the 
 white solid circles, centered at ($l$, $b$) $=$ (0.6676\arcdeg, $-$0.0226\arcdeg) with a radius of
 1.7\arcmin\ and at ($l$, $b$) $=$ (0.7394\arcdeg, $-0.087$\arcdeg) with a radius $=$1.5\arcmin, 
 respectively. The background region is shown by the white dashed ellipse centered at 
 ($l$, $b$) $=$ (0.6410\arcdeg, $-$0.1523\arcdeg) with the major and minor radii of 
 5.5\arcmin\ and 3.5\arcmin, respectively. The inner white dashed circle at ($l$, $b$) $=$ 
 (0.6885\arcdeg, $-$0.1523\arcdeg) with a radius $=$ 1.5\arcmin\ is excluded for the 
 background region (see text). 
 In this figure, 0.1\arcdeg\ (6\arcmin) corresponds to 50 light-years.
 The color scale is arbitrarily adjusted for comparison. 
 }
\label{fig:image}
\end{figure}

Figure~\ref{fig:image}a shows the X-ray images for 
the \ion{Fe}{1}-K$\alpha$ (6.4 keV) line of the Sgr B region in 2005, 
where bright diffuse sources are clearly detected \citep{Ko07b}. 
According to a previous work \citep{Ko07b}, we hereafter refer to the two sources as 
Sgr~B2 and M\,0.74$-$0.09. 
Figure~\ref{fig:image}b shows the same image taken four years later in 2009. 
Comparing the two images, 
we observe that the two bright sources have darkened in the four years, 
whereas there is no significant change in the brightness of the surrounding region. 

In addition, Figure~1c and 1d show the hard X-ray images of the Sgr B region 
in the 8--10~keV band for the 2005 and 2009 observations, respectively. 
As shown in Figure~\ref{fig:image}c, hard X-ray excesses at the positions of 
Sgr B2 and M\,0.74$-$0.09 in 2005 are observed. 
This is the first resolved hard X-ray image of Sgr B2 and M\,0.74$-$0.09,
although in the past X-ray observation, the hard X-ray excess near Sgr B2 has been found with 
an insufficient special resolution of 12\arcmin\ \citep{Te10}. 
On the other hand, the hard X-ray fluxes from these MCs are almost comparable to those 
of the surrounding areas and hence no significant hard X-ray excesses are found in the 
2009 data (Figure~\ref{fig:image}d). 
The significance levels of the hard X-rays from Sgr~B2 and M\,0.74$-$0.09 
with respect to the background region shown by the white dashed line are 
5 and 4$\sigma$ in 2005 and $<3\sigma$ in 2009, respectively.
These facts suggest that the hard X-rays from 
Sgr B2 and M\,0.74$-$0.09 decreased from 2005 to 2009 in correlation with 
the decrease in the \ion{Fe}{1}-K fluxes.

\subsection{Spectra}
\begin{figure}[tb]
 \epsscale{0.8}
 \plotone{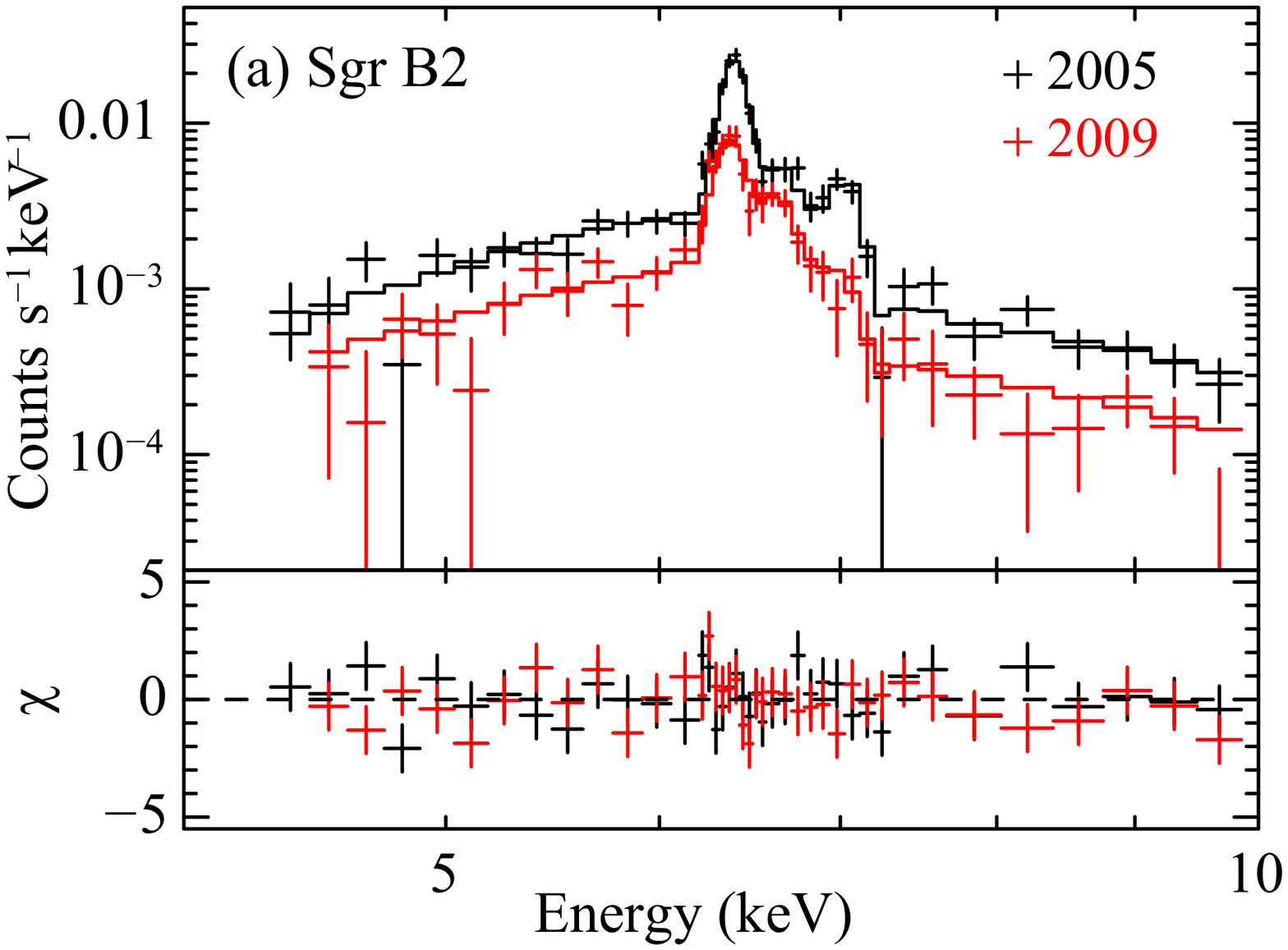}
 \epsscale{0.8}
 \plotone{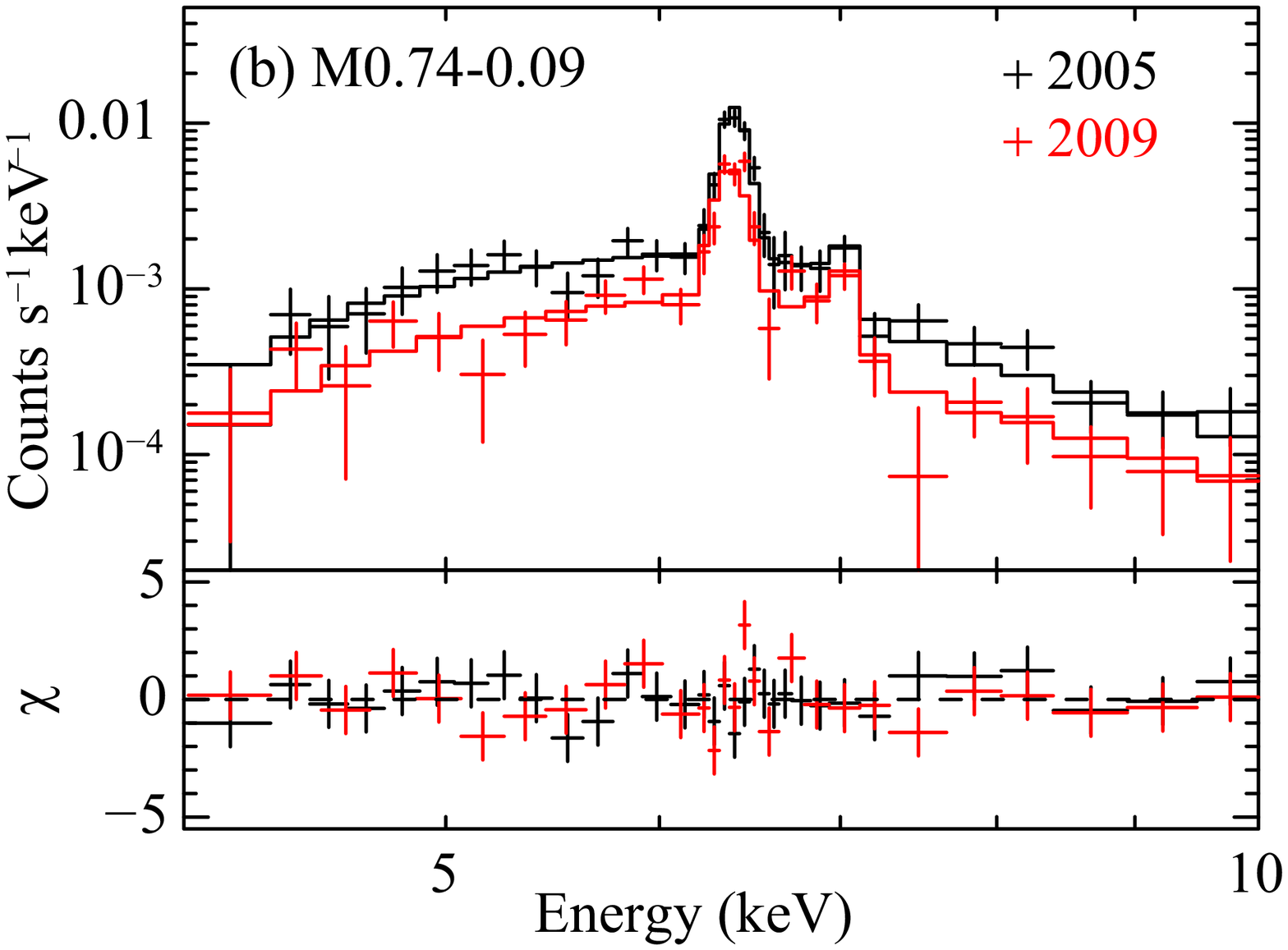}
 \caption{X-ray spectra of Sgr~B2 (a) and M\,0.74$-$0.09 (b). Spectra of the 2005 observation 
 and the 2009 observation are indicated with black and red crosses, respectively. 
 The best-fit models are shown with solid histograms.
 Residuals between the observed data and the models are shown in the bottom panels of (a) and (b).}
  \label{fig:spec}
\end{figure}

We investigate the X-ray spectra to study the time variability more quantitatively. 
The regions selected for the Sgr~B2 and M\,0.74$-$0.09 spectra are indicated with the white 
solid lines in Figure~\ref{fig:image}. The background region is also shown with  
white dashed lines, where the dashed small circle in which an MC might be emitting  
faint \ion{Fe}{1}-K is excluded. Figures~\ref{fig:spec}a and \ref{fig:spec}b 
are the background-subtracted spectra of Sgr B2 and M\,0.74$-$0.09, respectively. 
The spectra in 2005 and 2009 are shown by black and red crosses, respectively. 
As seen in both spectra, the fluxes of the \ion{Fe}{1}-K$\alpha$ lines and the hard 
continuum have simultaneously decreased from 2005 to 2009.

We fit the spectra with a phenomenological model of an absorbed power-law and the 
\ion{Fe}{1}-K$\alpha$ and \ion{Fe}{1}-K$\beta$ lines at 6.4 keV and 7.06 keV, respectively. 
A supernova remnants candidate G\,0.61+0.01 \citep{Ko07b} is located 4\arcmin\ away 
from Sgr B2 and may slightly contaminate the Sgr B2 spectrum. Therefore, we consider 
the contamination of the thermal plasma of G\,0.61+0.01 by adding a plasma model 
\citep{Ma98} with a fixed temperature of 37 million K and an iron abundance 
of \zfe$=$5.1 solar \citep{Ko07b}. 
The contamination is then found to be less than 10\%, and hence the error 
due to the possible uncertainty of flux contamination can be neglected.

The four spectra (Sgr B2 and M\,0.74$-$0.09 for 2005 and 2009) are simultaneously fitted with 
the free parameters of the absorption column density \nh, the iron abundance \zfe, 
the photon index \pwFP, and the fluxes of the continuum and the \ion{Fe}{1}-K lines. 
Among them, only \zfe\ and \pwFP\ are set to be common for all spectra. The fit is 
acceptable with chi$^2$/d.o.f. of 216/217 (null hypothesis probability $=0.6$). 
The best-fit common parameters of \zfe\ and \pwFP\ are 1.3$\pm$0.3 solar and 2.5$\pm$0.6 solar, 
respectively, and those of the other free parameters are listed in Table~\ref{tab:ana1}.

We find that 
the fluxes of the \ion{Fe}{1}-K$\alpha$ lines and the hard continuum band for 
Sgr~B2 decrease by 
a factor of 0.39$\pm$0.04 and 0.49$\pm$0.09 from 2005 to 2009, respectively. 
These variation factors are consistent within the 1$\sigma$ confidence level with the mean 
value of 0.41$\pm$0.04. The variation factors of 0.53$\pm$0.07 (\ion{Fe}{1}-K$\alpha$ lines) 
and 0.53$\pm$0.13 (continuum band) for M\,0.74$-$0.09 are also consistent with the mean 
value of 0.53$\pm$0.06.

\begin{table}[tb]
 \vspace{-0.2cm}
 \begin{center}
  \caption{Best-fit parameters\tablenotemark{a}.}
  \label{tab:ana1}
  \begin{tabular}{lllll}
   \tableline
   Source (Year)  & \nh   & \ion{Fe}{1} K$\alpha$ & \ion{Fe}{1} K$\beta$ & 5--10~keV \\
   unit          & $10^{23}$~cm$^{-2}$ & 10$^{-5}$\tablenotemark{b} & 10$^{-5}$\tablenotemark{b} & 
     10$^{-12}$\tablenotemark{c} \\
   \tableline
   Sgr~B2 (2005)          & 8.4$\pm$0.9 & 13.6$\pm$0.7 & 1.3$\pm$0.3 & 4.3$\pm$0.5 \\
   Sgr~B2 (2009)          & 8.8$\pm$1.4 &  5.2$\pm$0.5 & $<$0.4        & 2.1$\pm$0.3 \\
   M\,0.74$-$0.09 (2005)  & 5.7$\pm$0.4 &  5.1$\pm$0.4 & 0.4$\pm$0.2 & 1.7$\pm$0.2 \\
   M\,0.74$-$0.09 (2009)  & 6.5$\pm$1.2 &  2.7$\pm$0.3 & 0.5$\pm$0.2 & 0.9$\pm$0.2 \\
   \tableline
  \end{tabular}
 \vspace{-0.4cm}
  \tablenotetext{a}{The error ranges in this table are calculated at the 1$\sigma$ confidence level.}
  \tablenotetext{b}{ Observed flux in the unit of photons~s$^{-1}$~cm$^{-2}$.}
  \tablenotetext{c}{ Observed flux in the unit of ergs~s$^{-1}$~cm$^{-2}$.}
 \end{center}
\end{table}

\section{Discussion}\label{discussion}  
On the basis of the 2005 data,
we confirmed that the two MCs (Sgr~B2 and M\,0.74$-$0.09) exhibit intense 
\ion{Fe}{1}-K lines. In addition,
we resolved the hard continuum emission from Sgr B2 detected by $Granat$ and INTEGRAL
\citep{Su93, Re04} into emissions from two MCs.
The most important discovery is that the correlated decrease in the four years between 
the \ion{Fe}{1}-K lines and the hard continuum is synchronized in the two MCs,
although \cite{Te10} reported a correlated decrease in the entire Sgr~B region.
Because the data for these results 
are obtained with the same instrument (\textit{Suzaku}/XIS) and an identical pointing field, 
the possible systematic errors are minimized. 
The correlation of the variability between the two MCs
and the two components (\ion{Fe}{1}-K lines and hard continuum) indicate that they 
are of common origin.

As we have already discussed, the origin is likely to be X-ray irradiation by an external 
source (''X-ray echo''). Then, the equivalent width of the \ion{Fe}{1}-K$\alpha$ lines (EW$_{\rm 6.4keV}$) 
is expressed as 1.0$\times$(\zfe/1 solar)~keV \citep{Ta03, No10}. As the observed EW$_{\rm 6.4keV}$ 
is 1.0--1.5 keV, we derive the iron abundance as \zfe$=$1.0--1.5~solar. The absorption column density 
\nh\ of about 10$^{24}$ cm$^{-2}$ is significantly larger than the value of 10$^{23}$~cm$^{-2}$ for 
typical sources in the GC \citep{Mu04}, and hence, the major fraction would be owing to the absorption 
in the MCs. The iron absorption edge depth at 7.1~keV gives the iron abundance in the 
MCs of \zfe$= 1.3\pm0.3$ solar, which is consistent with that derived from EW$_{\rm 6.4keV}$. 
These facts firmly support the ''X-ray echo'' for the X-ray emission from Sgr B2 and M\,0.74$-$0.09: 
the fluorescent \ion{Fe}{1}-K line and the Thomson-scattered hard-continuum.

We found that the fluxes of Sgr B2 and M\,0.74$-$0.09 had reduced by factors of 
0.41$\pm$0.4 and 0.53$\pm$0.6 during the four years, respectively. 
This result shows that the decrease in \ion{Fe}{1}-K emission from 2000 to 2005 found by 
previous works \citep{Ko08, Inui09} had continued at least until 2009. 
\cite{Inui09} reported that the decrease rate of the \ion{Fe}{1}-K emission 
is $\sim$0.6 in five years (2000--2005). 
Thus, the X-ray fluxes had decreased by a factor of $\sim0.2$ during the decade.
In addition,
\cite{Te10} showed a decay of up to 0.4 in a hard X-ray flux from 2003 
to 2009, which is almost consistent with our result for the hard continuum.

\begin{figure}[tb]
 \epsscale{1.0}
 \plotone{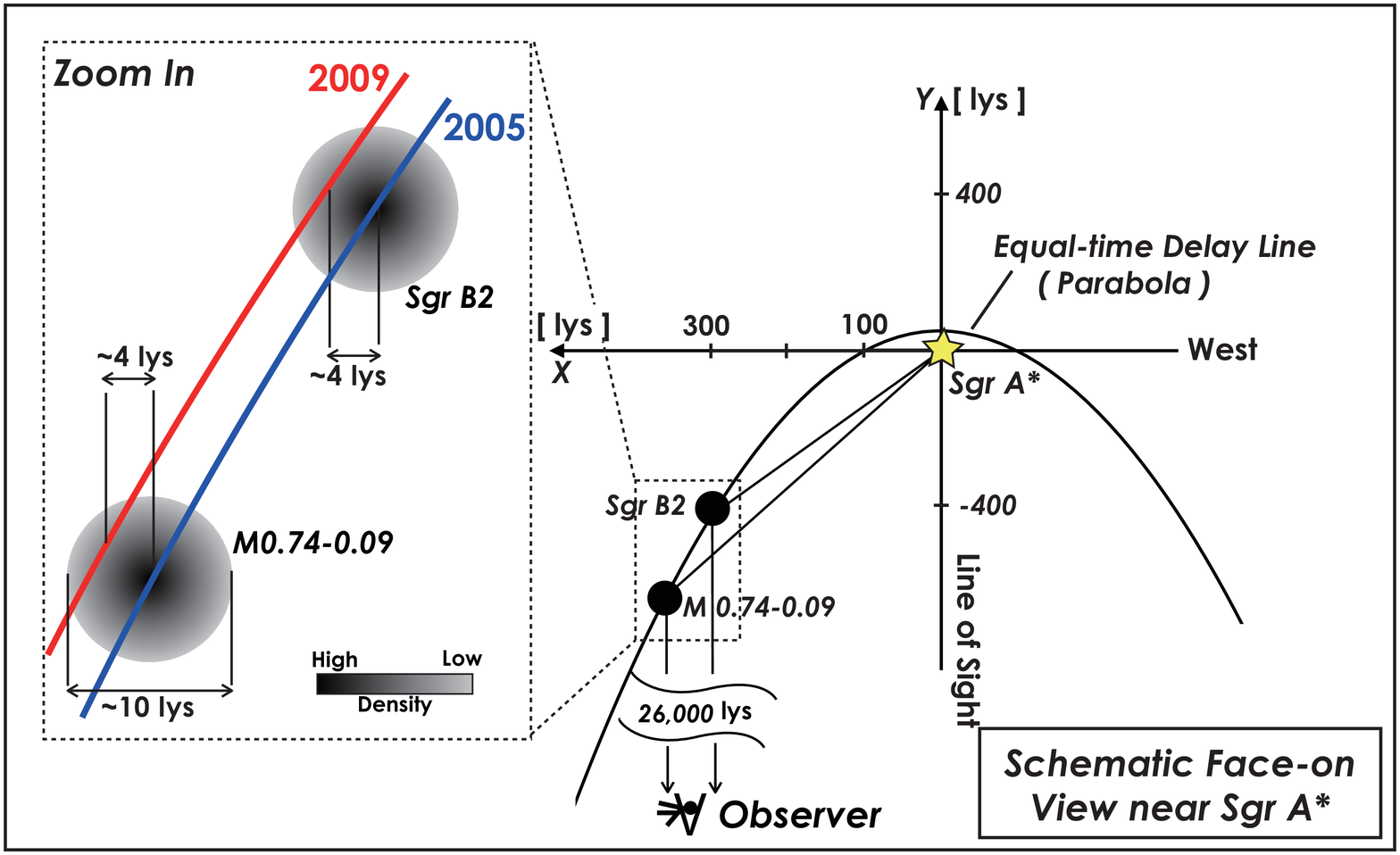}
 \caption{Schematic face-on view of the movement of the X-ray echoes in the molecular clouds. 
  The irradiating source, the super-massive black hole Sgr A*, is indicated with the 
  yellow star. The line-of-sight positions of Sgr~B2 and M\,0.74$-$0.09 are adopted from 
  the results of a previous work \citep{Reid09}. The solid parabola guides the eyes to see the 
  X-ray echoes from Sgr~B2 and M\,0.74$-$0.09 arriving at the observer at the same time 
  (equal-time delay line).The unit of distance is light-years (lys).}
  \label{fig:time-delay}
\end{figure}

The schematic view of the "X-ray echo" scenario is shown in Figure~\ref{fig:time-delay}. 
As the incident X-ray flare moves at light speed in the MCs, the echo spot also moves
at light speed. 
Because the density of the MCs is not uniform, the brightness of the X-ray echoes would be higher 
in high-density regions than in low-density regions. The X-ray echoes in 2005 
was produced in high-density regions and then moved into low-density regions in 2009. 
\cite{Po10} proposed that the flare lasted for a large duration of $\sim300$~years
(long flare). 
In the long flare scenario, is was suggested that the MCs in the Sgr B2 
region were illuminated by X-rays at the end of the long flare. 
On the other hand, 
\cite{Yu11} indicated that a single shorter flare can explain the X-ray light curve 
of Sgr B2.
We do not yet have enough data to determine the probable scenario.

The required luminosity (\lx) of the X-ray flare to produce the ''X-ray echo'' of the MCs depends 
on the time duration ($t$) of the flare, the distances ($d$) to the MCs, and the optical depths 
($\tau$) along the X-ray pass band in the MCs. In particular, 
\lx\ also depends on the density profiles and the sizes of MCs. 
The MC sizes may be $\sim$10 light years, almost 
equal to those in the selected regions shown 
in Figure~\ref{fig:image}.
Using the observed 
\nh\ of about 10$^{24}$~cm$^{-2}$, $\tau$ is estimated to be about 0.1. 
Then, for the two cases of long flare $t > 10$~years and short flare $t < 10$~years,
we can estimate \lx\ from \citet{Su98} as follows:
\begin{eqnarray}
\lx = \left\{
\begin{array}{ll}
4\times 10^{39}\ \left( \frac{t}{\rm 10 year} \right)^{-1}
       \left( \frac{d}{\rm 500~ly} \right)^2
        {\rm erg~s}^{-1} & 
         (t \leq 10{\rm ~ year}) \\
4\times 10^{39}\ \left( \frac{d}{\rm 500~ly} \right)^2
        {\rm erg~s}^{-1} &
         (t > 10{\rm ~ year}.) \\
\end{array} \right.
\label{eq:reqiredLumin}
\end{eqnarray}

Using radio observations,
\cite{Mu00} indicated that the X-ray peak shifted from the MC core
to the east direction by $\sim1.'2$; therefore, the X-ray flare source should be 
located on the west side of Sgr B2.
Moreover, the minimum value of the required luminosity is given when the 
flare-source is near the west-rim of Sgr B2. 
In this case, the distance $d$ to the other MC, M\,0.74$-$0.09 is about 
0.1\arcdeg or 50 light-years (Figure~\ref{fig:image}). 
The minimum value of the required luminosity is \lx $= 4 \times 10^{37}$~erg~s$^{-1}$. 
As Equation~(\ref{eq:reqiredLumin}) is derived from simple assumptions and approximations, 
some uncertainties exist in \lx $= 4\times10^{37}$~erg~s$^{-1}$, but the errors should not 
be larger than one order of magnitude. 

The most conservative value of \lx\ is $4\times10^{36}$~erg~s$^{-1}$, which is almost 
equal to that of the brightest X-ray sources in our Galaxy. 
Because we did not find such stable sources 
in the GC, one possible candidate is a transient X-ray binary system. 
X-ray monitoring in the GC has been conducted 
since almost 10 years with the \textit{INTEGRAL} satellite \citep{Ku07}. All bright 
transient sources exhibited peak luminosities lower than a few times of 10$^{37}$~erg~s$^{-1}$ 
\citep{Sa02, We03}. Moreover, the flare durations ($t$) are shorter than 0.1 years \citep{Sa02, We03}. 
The average luminosity of the transient sources over 10 years is a few times of 
$10^{35}$~erg~s$^{-1}$,
which is 10 times lower than the minimum required luminosity by the most conservative 
estimation. 

These facts indicate that the transient binary system is an unlikely candidate for the irradiation source. 
Therefore, we conclude that the large-flare of Sgr A*, the super-massive black hole,
is the irradiation source. 
The large flare with a luminosity of more than $4\times10^{39}$~erg~s$^{-1}$ might have 
occurred a few hundred years ago. 
At present, the X-ray luminosity of Sgr A* is about 
$10^{34}$~erg~s$^{-1}$ and it exhibits numerous flares. However, the flare scales are relatively small 
and their luminosity can at most reach 180 times that of the
quiescent level \citep{Po03}. The large flare predicted in this study, 
which is more than 10$^4$ times the largest flare observed until now, would be new 
evidence for past high activity of the super-massive black hole at the center of our Galaxy.

\vspace{-0.5cm}
\acknowledgments
This work is supported by the Grant-in-Aid for the Global COE Program 
''The Next Generation of Physics, Spun from Universality and Emergence''
from the Ministry of Education, Culture, Sports, Science and Technology 
(MEXT) of Japan. This work is also supported by Grant-in-Aids from the 
Ministry of Education, Culture, Sports, Science and Technology (MEXT) of 
Japan, Scientific Research A, No. 18204015 (KK), and Scientific Research B, 
No. 20340043 (TT). 
MN and SGR are supported by JSPS Research Fellowship for Young Scientists.

\end{document}